\documentclass[twocolumn,showpacs,preprintnumbers,amsmath,amssymb]{revtex4}
\usepackage[utf8]{inputenc}
\usepackage{graphicx}
\usepackage{dcolumn}
\usepackage{bm}
\usepackage{tikz}
\usetikzlibrary{arrows,automata}
\usepackage{graphicx}
\usepackage{dcolumn}

\usepackage{amssymb}
\usepackage{amsmath}

\usepackage{booktabs}
\usepackage{mathtools}
\usepackage{array}

\begin{document}

\preprint{APS/123-QED}

\title{Exact performance of the five-qubit code with coherent errors }


\author{Chaobin Liu}
\email{cliu@bowiestate.edu}
\affiliation{Department of Mathematics, Bowie State University, Bowie, Maryland 20715, USA}

\begin{abstract}

To well understand the behavior of quantum error correction codes (QECC) in noise processes, we need to obtain explicit coding maps for QECC. Due to extraordinary amount of computational labor that they entails, explicit coding maps are a little known. Indeed this is even true for one of the most commonly considered quantum codes-the five-qubit code, also known as the smallest perfect code that permits corrections of generic single-qubit errors. With direct but complicated computation, we obtain explicit process matrix of the coding maps with a unital error channel for the five-qubit code. The process matrix allows us to conduct exact analysis on the performance of the quantum code. We prove that the code can correct a generic error in the sense that under repeated concatenation of the coding map with itself, the code does not make any assumption about the error model other than it being weak and thus can remove the error(it can transform/take the error channel to the identity channel if the error is sufficiently small.). We focus on the examination of some coherent error models (non diagonal channels) studied in recent literatures.  We numerically derive a lower bound on threshold of the convergence for the code. Furthermore, we analytically show how the code affects the average gate infidelity and diamond distance of the error channels. Explicit formulas of the two measurements for both pre-error channel and post-error channel are derived, and we then analyze the logical error rates of the aforesaid quantum code. Our findings tighten the upper bounds on diamond distance of the noise channel after error corrections obtained in literature.

\end{abstract}

\maketitle

\section{Introduction}

Quantum error correction (QEC) is one of the most vital components of quantum information science \cite{NC2000}. It was well known from the very start of this exciting field that the delicate coherent quantum system would be particularly fragile in the presence of noise. The fragility of coherent quantum systems would be a major obstacle to the development of large scale quantum computers. The introduction of quantum error correction in the middle of 1990s \cite{Shor1995, Steane1996a, Bennett1996, Laflamme1996} revealed that active techniques could be used to prevent noise in the underlying systems from causing logical errors. Our understanding has advanced significantly since then \cite{G1997, FM2012, DNM2013, T2015, Brun2019}.

A large research effort has been recently made to examine the performance of QEC at the logical level under various noises, and to realize quantum codes in lab. To quantify the error magnitude at the logical (physical) levels, various metrics have been utilized such as threshold, the average gate infidelity, or diamond distance of the noise channel to the identity channel. \cite{GS2016} studied the logical error rate of the Steane code by performing full-density-matrix simulations of an error correction step against some incoherent and coherent noise channels. \cite{DP2017} used tensor-network simulation to study the threshold and the subthreshold behavior of the amplitude-damping and systematic rotation channels for the surface code. \cite{BEKP2017} simulated quantum error correction protocols based on the surface code and obtained the first error threshold estimates for certain models of coherent noise. The decay rate was analyzed for the specific noise model (determined by the rotation along Pauli $X$ and a stochastic bit flip) in the repetition code \cite{GD2018}. \cite{BW2018} provided a bound of the logical error in terms of the average gate infidelity for codes correcting a general noise channel. \cite{IP2020} derived an inequality which relates the diamond distance of the logical channel from the identity to the average infidelity of the logical noise channel for the toric code. \cite{CXB2020} showed that some of the coherence of the noise channel can be used to improve its logical fidelity by using Pauli conjugation in quantum error correction. \cite{BW2021} proposed efficient method of computing logical noise in quantum error correcting codes. \cite{GY2022} experimentally realized five-qubit code and demonstrated each key aspect of the code (such as implementing logical Pauli operators). 

Notably the coding map method introduced in \cite{RDM2002} has proven fruitful in the study of performance of quantum correction codes.  By encoding each physical qubit of an error correcting code into another code, and the procedure can be repeated recursively, a concatenated code is then formed. \cite{RDM2002} showed how to efficiently compute threshold values for several specified concatenated codes (the five-qubit code, the Steane's seven-qubit code and the Shor's nine-qubit code) under a fixed decoder when each qubit is afflicted by diagonal noise channel. A couple of groups of researchers have followed the coding map methods to evaluate effective noise channels for the encoded qubits after error correction. \cite{FKSS2006} considered the coding map as a discrete-time dynamical system on the entire space of noise channels and extended the result of \cite{RDM2002}, in the case of diagonal channels, they proved that any code with distance at least three corrects (in the infinite concatenation limit) an open set of errors. In the language of dynamical system, the identity channel is locally attracting. \cite{CW2017} continued to investigate the the performance of the commonly considered concatenated codes including surface-17 code, they used hard decoding algorithm for optimizing thresholds (the set of correctable errors) under certain selected non-diagonal noise channels. Using the coding map as a tool, \cite{HDF2019} obtained an upper bound on the diamond norm for a wide range of codes against a specified model of coherent noise.

The coding map of a quantum code synthesizes encoding, noise, syndrome measurement, recovery and decoding to form the so-called process matrix, by which one could obtain the effective logical channel from the physical noise channel. The process matrix allows one to conduct exact analysis on the performance of the quantum code. For a code encoding one logical qubit into $n$ physical qubits, the number of syndromes grows as $2^n$ making it rapidly become unmanageable to keep track of all of them for lager codes, and therefore explicit process matrices of coding maps for quantum codes under a generic noise model are little known (this is even true for the smallest possible nontrivial quantum code-the five-qubit code). To our best knowledge, no explicit process matrix of a quantum code for a general noise channel is available in literature. In this paper, we obtain explicit process matrix of the coding maps under any unital error channel for the five-qubit code.  We then prove that the concatenated code can correct an open set of errors. For a specified coherent error model studied recently in literature, explicit formulas of the two measurements (average gate infidelity and diamond distance)
for both physical error channel and logical error channel are derived, and we then analyze the logical error
rates of the aforesaid quantum code. Our findings tighten the upper bounds on diamond distance
of the noise channel after error corrections obtained in literature \cite{HDF2019}.

The paper is structured as follows. In Section II we review the coding map method. In Section III, we present one of our main results about the convergence of the concatenated five-qubit coding map with the symmetric decoder. The explicit process matrix for the coding map is  given in Appendix A. We also conduct technical discussion on lower bounds to threshold values of the attraction of the identity channel for a specific noise model proposed by previous author, the numerical tests and analysis can be found in Appendix B.  In section IV, we analytically show how the code affects the average gate infidelity and diamond distance of the error channels. Explicit formulas of the two measurements for both pre-error channel and post-error channel are derived. Details of the deriving the diamond distance is provided in Appendix C.

\section{Coding map method review}

We present the basic framework laid out and review relevant results from \cite{RDM2002, FKSS2006}, which should be refereed to for more details. Quantum states are represented by their density matrices.

The error correction process consists of three consecutive actions: encoding $\mathcal{E}$, noise $\mathcal{N}$, and decoding $\mathcal{D}$. Each action is modeled as a {\it quantum channel}.

Encoding $\mathcal{E}$ takes an initial logical qubit state $\rho_0$ to the initial register state $\rho(0)$, which is assumed to evolve according to a discrete-time operation $\mathcal{N}$ taking $\rho(0)$ into a final register state $\rho(t)=\mathcal{N}(\rho(0))$. Finally, decoding $\mathcal{D}$ takes $\rho(t)$ to the final logical qubit state $\rho_f$. The map 

\begin{eqnarray}
\mathcal{G}=\mathcal{D}\circ\mathcal{N}\circ\mathcal{E}: \rho_0\rightarrow \rho_f
\end{eqnarray}

\noindent describes the effective dynamics of the encoded information resulting from the physical dynamics of $\mathcal{N}$ and is called the {\it effective (logical) channel}.

We consider noise models $\mathcal{N}$ on qubits consisting of uncorrelated noise $\mathcal{N}^{(1)}$ on each single physical qubit, so

\begin{eqnarray}
\overset{\hskip 0.4 in n\, \mathrm{copies}}{\mathcal{N}=\overbrace{\mathcal{N}^{(1)}\otimes...\otimes \mathcal{N}^{(1)}}}
\end{eqnarray}

\noindent Given an $n$ qubit quantum error correcting code $\mathcal{C}$ with encoding operation $\mathcal{E}$ and decoding operation $\mathcal{D}$ , the map taking
the single qubit (physical) noise $\mathcal{N}^{(1)}$ to the effective channel $\mathcal{G}$

\begin{eqnarray}
\Omega^{\mathcal{C}}:\, \mathcal{N}^{(1)}\rightarrow \mathcal{G}=\mathcal{D}\circ (\mathcal{N}^{(1)})^{\otimes n}\circ \mathcal{E}
\end{eqnarray}

\noindent is called the coding map of $\mathcal{C}$.

Let $\mathcal{C}$ be a stabilizer code given by stabilizer $\mathcal{S}=\{S_k\}\subset \pm\{I, X, Y, Z\}^{\otimes n}$, storing one qubit in an $n-qubit$ register. The stabilizer $\mathcal{S}$ defines the codespace, and the logical operators $\bar{I}, \bar{X}, \bar{Y},\bar{Z}\subset \pm\{I, X, Y, Z\}^{\otimes n}$ determine the particular basis of codewords $|\bar{0}\rangle, |\bar{1}\rangle$. Let $E_\sigma$ denote $\frac{1}{2}\mathcal{E}[\sigma]$, then the encoding operation $\mathcal{E}$ can be completely characterized by $E_\sigma$ operators, which can be expressed as

\begin{eqnarray}
\!\!\!E_\sigma
\!\!&=&\frac{1}{2|S|}\sum_k S_k\bar{\sigma} \nonumber\\
\!\!&=&\!\!\sum_{\mu_i \in \{I,X,Y,Z\}}\alpha_{\{\mu_i\}}^{\sigma}(\frac{1}{2}\mu_1)\otimes...\otimes(\frac{1}{2}\mu_n).
\end{eqnarray}

\noindent The decoding operation $\mathcal{D}$ can be completely characterized by $D_\sigma$ operators

\begin{eqnarray}
\!\!\!D_\sigma
\!\!&=&2\sum_j R_j^{\dagger}E_\sigma R_j \nonumber\\
\!\!&=&\frac{1}{|S|}\sum_{j,k} R_j^{\dagger}S_k\bar{\sigma} R_j\nonumber\\
\!\!&=&\!\!\sum_{\nu_i \in \{I,X,Y,Z\}}\beta_{\{\nu_i\}}^{\sigma}(\nu_1)\otimes...\otimes(\nu_n)
\end{eqnarray}

\noindent where $R_j$ is a recovery operator, which will be explained later on.

The density matrix of one qubit state can be expanded in the standard basis $\mathcal{P}=\{P_1,P_2, P_3, P_4\}=\{I, X, Y, Z\}$, and represented as a four-dimensional real vector. A noise channel
can then be represented as a $4\times 4$ matrix called the Pauli-Liouville representation (which has matrix elements $(\mathcal{N}^1)_{ij}\doteq \frac{1}{2}\mathrm{Tr}[P_i\mathcal{N}^1(P_j)]$


\begin{equation}
\mathcal{N}^{(1)}= \left[\begin{array}{cccc}
1& 0 &0&0 \\
N_{XI}      & N_{XX} & N_{XY}&N_{XZ}\\
N_{YI} & N_{YX}& N_{YY}&N_{YZ}\\
N_{ZI}&N_{ZX}&N_{ZY}&N_{ZZ}
\end{array}\right]
\end{equation}

Zeroes in the first row are due to trace preservation. If a noise channel is unital ( $\mathcal{N}^{(1)}(\mathbb{I}_2)=\mathbb{I}_2$), then it is evident that its matrix representation becomes

\begin{equation}
\mathcal{N}^{(1)}= \left[\begin{array}{cccc}
1& 0 &0&0 \\
0      & N_{XX} & N_{XY}&N_{XZ}\\
0& N_{YX}& N_{YY}&N_{YZ}\\
0&N_{ZX}&N_{ZY}&N_{ZZ}
\end{array}\right] \label{unitalNC}
\end{equation}

For an arbitrary $n$ qubit code $\mathcal{C}$, the entries of the matrix $\mathcal{G}=\Omega^{\mathcal{C}}(\mathcal{N}^{(1)})$ can be calculated by

\begin{equation}
\mathcal{G}_{\sigma \sigma^{\prime}}=\sum_{\{\nu_i\}}\sum_{\{\mu_i\}}\beta_{\{\nu_i\}}^{\sigma}\alpha_{\{\mu_i\}}^{\sigma^{\prime}}\Pi_{i=1}^{n} N^{(1)}_{\nu_i\mu_i} \label{echannel}
\end{equation}
where $\{\mu_i\}$ and $\{\nu_i\}$ run over $\{I, X, Y, Z\}^{\otimes n}$.

We point out that the decoding operation is defined to be $\mathcal{D}=\mathcal{E}^{\dagger}\circ \mathcal{R}$ where $\mathcal{R}$ includes the measurement update and recovery map. To build intuition for $\mathcal{D}$, we note that for most codes considered in the literature (and 
the specific code called $k=1$ code considered later in this paper) the
recovery procedure is given in a particular form. The $n$ dimensional Pauli group is $\mathcal{P}_n=\{\pm 1,\pm i\}\otimes \{I, X, Y, Z\}^{\otimes n}$. Suppose we have a stabilizer code $\mathcal{C}$ that encodes $k$ qubits into $n$ qubits. Its stabilizer $\mathcal{S}$  is an Abelian subgroup of $\mathcal{P}_n$ with $n-k$ generators $g_i$. There are $2^{n-k}$ orthogonal $2^k$-dimensional subspaces of the register state space.  Each of the subspaces is the intersection of eigenspaces of the generators $g_i$, which can be given by projector $P_\beta=\frac{1}{2^{n-k}}\Pi_{j=1}^{n-k}(I+(-1)^{\beta_j}g_j)$ where $\beta=(\beta_1, \beta_2,..., \beta_{n-k})\in \mathbb{Z}_2^{n-k}$. It is noted that when $\beta=(1,1,...,1)$, the subspace projected by $P_{\beta}$ is the code space denoted by $\mathcal{C}_{\mathcal{S}}$. When the register state evolves with a correctable error, the code space $\mathcal{C}_{\mathcal{S}}$ is then transformed into the subspace projected by some $P_\beta$. After a syndrome measurement is made, the aforesaid $\beta$ would be detected, the corresponding recovery operator $R_\beta$ acts on the register, unitarily mapping the subspace projected by $P_\beta$ back to the code space $\mathcal{C}_{\mathcal{S}}$. Therefore the decoding operation can
be expressed as a sum over all syndromes, $\mathcal{D}=\sum_\beta \mathcal{E}^{\dagger}\circ R_\beta\circ P_\beta$.

\section{Open set of correctable unital non-diagonal errors}

 \subsection{Some basics of the theory of dynamical systems}

We present some basics about the theory of dynamical systems which will be used to analyze concatenated quantum codes. 

A dynamical system can be described by a vector-valued map $f: S\rightarrow S$ where $S$ is a measurable subset of $\mathbb{R}^k$, and $f$ is a differentiable function. The total derivative (or the Jacobi matrix ) of $f$ at point $p$ is given by $\mathrm{D}f(p)=(\frac{\partial f_i}{\partial x_j}|_{x=p})_{k\times k}$ where $f_i$ is the $i$-th component function of $f$. A norm of $\mathrm{D}f(p)$ as an operator on $\mathbb{R}^k$ can be defined as $||\mathrm{D}f(p)||=\underset{||v||=1}{\mathrm{max}} ||\mathrm{D}f(p)v||$, this version of norm is known as the matrix 2-norm, and can be also calculated by $||A||_2=\sqrt{\rho(A^{\star}A)}$ where $\rho(A^{\star}A)$ is the largest eigenvalue of $A^{\star}A$ and $A^{\star}$ is the complex conjugate of $A$.

A basic question of dynamical systems is to ask:  what happens to the system in the long run? Precisely speaking, where does $f^n(p)$ converge as $n$ goes to $\infty$ for a given point $p$ (representing an initial state of the system) ? Here $f^n=f \circ \cdots \circ f$ ($n$ times of the composition function of $f$ with itself). Perhaps an interesting and important point/state to which $f^n$ converges should be the {\it fixed point} of $f$ (which means $f(p)=p$), an even more interesting and important point should be the so-called {\it locally attracting} point,  which means that a point $p$ is fixed and if there exists a neighborhood $V$ of $p$ such that for every point $x\in V$, $f^n(x)\rightarrow p$ as $n\rightarrow \infty$. The largest such set $V$ is called the {\it basin of attraction} of the fixed point $p$. A standard criterion for a fixed point to be locally attracting is given as follows (see \cite{FKSS2006} and the references therein.):

Lemma 1:\,\, Suppose $U\subset \mathbb{R}^k$ is open, $f:U\rightarrow \mathbb{R}^k$ is a $C^1$ map. $p\in U$ is a fixed point of $f$, and $||\mathrm{D}f(p)||<1$. Then $p$ is locally attracting.

\subsection{A coherent error channel}
We consider noise models $\mathcal{N}$ on qubits consisting of uncorrelated noise $\mathcal{N}^{(1)}$ on each single physical qubit, so

\begin{eqnarray}
\overset{\hskip 0.4 in n\, \mathrm{copies}}{\mathcal{N}=\overbrace{\mathcal{N}^{(1)}\otimes...\otimes \mathcal{N}^{(1)}}}
\end{eqnarray}

We will focus on analysis of convergence of the codes for the single-qubit error model considered in \cite{GD2018}, as shown below:

\begin{eqnarray}
\!\!\!\mathcal{N}^{(1)}[\rho]
\!\!&=&(1-q)e^{-i\epsilon X/2}\rho e^{i\epsilon X/2}+q Xe^{-i\epsilon X/2}\rho e^{i\epsilon X/2}X \nonumber\\
\!\!&=&\!\!\Lambda_{\epsilon}\circ\Lambda_q[\rho] \label{noiseX}
\end{eqnarray}

\noindent where $q$ is the probability of a stochastic bit-flip and $\epsilon$ is the angle of a small rotation error that is constant in time. Eq. (\ref{noiseX}) is considered to describe the composition of a coherent process, $\Lambda_{\epsilon}$, and an incoherent process, $\Lambda_q$.

Since $\mathcal{N}^{(1)}$ is a unital channel, its Pauli-Liouville representation (see Eq. (\ref{unitalNC}) is given below:

\begin{equation}
\mathcal{N}^{(1)}= \left[\begin{array}{cccc}
1& 0 &0&0 \\
0      & 1 & 0&0\\
0 & 0& (1-2q)\cos(\epsilon)&(2q-1)\sin(\epsilon)\\
0&0&(1-2q)\sin(\epsilon)&(1-2q)\cos(\epsilon)
\end{array}\right] \label{noiseM}
\end{equation}

We will concentrate on the coherent error channel defined by Eq. (\ref{noiseX}). It turns out that the characterization of the asymptotic properties of the coding map for a concatenated code is determined by examining the long-time behavior of the dynamical system 
$\Omega^C: \Delta\rightarrow \Delta$ where $\Delta$ is a subset of $\mathbb{R}^k$.

\subsection{The five-bit code with the symmetric decoder}

According to the process matrix of coding map for 5-qubit code we obtained (see Appendix A), for an input $M$, the output through the coding map $\Omega^{\mathrm{Five}}$ is given as follows:

\begin{equation}
\Omega^{\mathrm{Five}}(M)= \left[\begin{array}{cccc}
1& 0 &0&0 \\
0      & g(x,y,z)& 0&0\\
0 & 0& g(y,z,x)&h(v)\\
0&0&h(u)&g(z,x,y)
\end{array}\right] \label{postnoise}
\end{equation}

with \begin{equation}
M= \left[\begin{array}{cccc}
1& 0 &0&0 \\
0      & x& 0&0\\
0 & 0& y&v\\
0&0&u&z
\end{array}\right] \label{noiseM1},
\end{equation}

\noindent $g(x,y,z)=-\frac{x}{4}(x^4-5y^2-5z^2+5y^2z^2)$ and $h(w)=-\frac{w^5}{4}$.

As shown above, the five-bit coding map takes noise model given by Eq. (\ref{noiseM1}) to an effective channel $\mathcal{G}$ with the same pattern, thus we can treat the coding map $\Omega^{\mathrm{Five}}$ as a vector-valued map expressed below:

 \begin{eqnarray}
(x,y,z,u,v )\mapsto (g(x,y,z), g(y,z,x),g(z,x,y),h(u),h(v)) \label{dsfivebit}
\end{eqnarray}

\noindent It is noted that $(1,1,1,0,0)$ is a fixed point (as expected) because $\Omega^{\mathrm{Five}}(1,1,1,0,0)=(1,1,1,0,0)$. Using the partial derivatives of functions $g$ and $h$ with respect to variables $x,y,z,u$, and $v$, we can find that the total derivative $\mathrm{D}\Omega^{\mathrm{Five}}(1,1,1,0,0)=0$, which immediately implies the norm $||\mathrm{D}\Omega^{\mathrm{Five}}(1,1,1,0,0)||=0$.
By Lemma 1, $(1,1,1,0,0)$ is a locally attracting point of the coding map $\Omega^{\mathrm{Five}}$, therefore we conclude that the concatenated five-bit code can take a noise channel to the identity channel when the noise channel is sufficiently close to the identity channel.

Now we consider the noise channel in particular given by Eq.(\ref{noiseM}), denoted by $(x_0,y_0,z_0,u_0,v_0)$,  as the initial state of the dynamical system $\Omega^{\mathrm{Five}}$, and expressed below:

$\left\{
\begin{array}{l}
x_0=1\\
y_0=(1-2q)\cos(\epsilon)\\
z_0=(1-2q)\cos(\epsilon)\\
u_0=(1-2q)\sin(\epsilon)\\
v_0=(2q-1)\sin(\epsilon)
\end{array}\right.$

\noindent It can be seen that $(x_0,y_0,z_0,u_0,v_0)\rightarrow (1,1,1,0,0)$ as $q,\epsilon\rightarrow 0$. Therefore, when both $q$ and $\epsilon$ are sufficiently small, the noise channel $(x_0,y_0,z_0,u_0,v_0)$ falls in the basin of attraction of the identity channel $(1,1,1,0,0)$, then the dynamical system $\Omega^{\mathrm{Five}}$ will take the noise channel to the identity channel.

\vskip 0.2in

\noindent {\bf Technical discussion on lower bounds to threshold values of the attraction of the identity channel}\,\, We will figure out how small both $q$ and $\epsilon$ needed to be so that the error channel falls into the basin of attraction of the identity channel, and thus can be corrected. According to the five-bit coding map given by Eq.(\ref{dsfivebit}), each of the last two components of the map $h(w)=-\frac{w^5}{4}$ is solely dependent on a single variable (input) $w$ , and $h^n(w)\rightarrow 0$ as long as $|w|<1$. On the other hand, when $y=z$, the first component of $\Omega^{\mathrm{Five}}$ is $g(x,y,z)=-\frac{x}{4}(x^4+5y^4-10y^2)$,  and the second and third components of $\Omega^{\mathrm{Five}}$ become identical, in other words, $g(y,z,x)=g(z,x,y)=-\frac{y}{4}(y^4+5x^2y^2-5x^2-5y^2)$. Therefore, the convergence of the concatenated five-bit code actually amounts to the convergence of the dynamical system defined by $f=(f_1, f_2): (x,y)\mapsto (f_1(x,y), f_2(x,y))$ where $f_1(x,y)=-\frac{x}{4}(x^4+5y^4-10y^2)$, $f_2(x,y)=-\frac{y}{4}(y^4+5x^2y^2-5x^2-5y^2)$, and $(x,y)\in \Delta\subset \mathbb{R}^2$. In this present scenario, the point (state) $(1,1)$ is a locally attracting point of $f$ as $\mathrm{D}f(1,1)=0$. By Lemma 1, $||\mathrm{D}f(x,y||$ allows us to determine a neighborhood of $(1,1)$ that is entirely contained in the basin of $(1,1)$. Since the computation involved becomes unmanageable, we may have to use numerical approach to do the job, we here report the outcome we can achieve: via massive numerical tests and analysis (see Appendix B), we conclude that $||\mathrm{D}f(x,y)||<.995$ when $(x,y)\in \mathrm{B}((1,1), 0.072)$ where $\mathrm{B}(z,r)$ denotes the open ball of radius $r$ centered at $z$. This implies that when $1-(1-2q)\cos(\epsilon)<0.072$, the error channel given by Eq. (\ref{noiseM}) falls into the basin of the identity channel, and thus the concatenated five-bit code map takes the error channel to the identity channel and the error is corrected. It is noteworthy that when $q=0$, the purely coherent error channel can be corrected if the angle of  rotation error $\epsilon<\sqrt{0.144}\approx 0.379$. Remarkably, some recent studies such as  \cite{CW2017, HDF2019} provided similar results for different error models.

\vskip .2in

Finally, we conclude our discussion of the convergence of the concatenated five-qubit coding map by showing a general result on an arbitrary unital channel.

Theorem\,\, If $C$ is the five-qubit code with the symmetric decoder and the error channel $\mathcal{N}^{(1)}$ is unital (i.e. $\mathcal{N}^{(1)}(\mathbb{I}_2)=\mathbb{I}_2$ ), then the identity channel $\mathbb{I}_4$ is a locally attracting point of the coding map $\Omega^{C}$. 

Proof\,\,  For clarity the submatrix obtained by deleting the first row and first column of $\mathcal{N}^{(1)}$ given by Eq. (\ref{unitalNC}) is represented by $(N_{XX}, N_{YY}, N_{ZZ}, \cdots, N_{ZY})$, then $(1,1,1,0, \cdots, 0)$ represents the $3 \times 3$ identity matrix. By Lemma 1, it suffices to show that $\mathrm{D}\Omega^{C}(1,1,1,0,\cdots, 0)=0$. We will here only show the entry in the first row and first column of the $9 \times 9 $ matrix $\mathrm{D}\Omega^{C}(1,1,1,0,\cdots, 0)$ is zero, the proof of rest entries being zero can be given in a similar way. Based on the coding map formulas obtained in Appendix A, $\frac{\partial G_{XX}}{\partial N_{XX}}|_{(1,1,1,0,\cdots, 0)}=\frac{\partial h_x}{\partial x_1}|_{(1,1,1,0,\cdots, 0)}=-\frac{1}{4}(x_1^4+5x_2^2x_3^2+5y_1^2z_1^2+5y_2^2z_3^2)-\frac{x_1}{4}(4x_1^3)+\frac{5}{4}(y_2^2+z_3^2)|_{(1,1,1,0,\cdots, 0)}=0$.
 
\vskip 0.1in
It is noted that this theorem implies that the five-qubit code does not make any other assumption about the noise model other than that it should be weak and thus can be removed by applying the concatenated coding map repeatedly. It should be pointed out that \cite{FKSS2006} proved a similar result for a special case where the error channel is diagonal, and the proof used a different approach.

\vskip 0.2in

\section{How the quantum code affects the noise strength}

One usually wants to quantify the strength of errors $\mathcal{E}$ in quantum operations. There are two commonly used measures. One measure is the
average gate infidelity $r(\mathcal{E})$ ( to the identity, the perfect gate), 

\begin{eqnarray}
r(\mathcal{E})=1-\int \mathrm{d}\psi \langle \psi|\mathcal{E}(|\psi \rangle \langle \psi|)|\psi\rangle
\end{eqnarray}

\noindent which is considered to capture average-case behavior for a single use of the gate. This quantity can be estimated efficiently in randomized benchmarking experiments \cite{KLR2008, MGE2011}, or can be precisely calculated by the formula below \cite{N2002, HHH1999, KLDF2015}. 

Proposition 1.\,\, Let $\mathcal{E}$ be a completely positive (but not necessarily trace preserving) map with Pauli-Liouville representation $\mathrm{L}(\mathcal{E})$. Then,

\begin{eqnarray}
\mathrm{F_{avg}}(\mathcal{E})=\frac{\mathrm{Tr}[\mathrm{L}(\mathcal{E})]+\mathrm{Tr}[\mathcal{E}(I)]}{d(d+1)}, \label{fidelity}
\end{eqnarray}

\noindent where $\mathrm{F_{avg}}(\mathcal{E})=1-r(\mathcal{E})$ is the average fidelity and $d$ is the system size.

The other measure is called the diamond distance $\mathrm{D}_\diamond(\mathcal{E})$ ( to the identity), 

\begin{eqnarray}
\mathrm{D}_\diamond(\mathcal{E})= \frac{1}{2}||\mathcal{E}-\mathcal{I}||_\diamond = \frac{1}{2}\underset{\rho}{\mathrm{sup}}||(\mathcal{E}\otimes \mathcal{I})(\rho)-\rho||_1
\end{eqnarray}

\noindent where $||A||_1=\sqrt{\mathrm{Tr}(A^\dagger A)}$ and the supremum is over all pure states \cite{K1997}. This measure describes the worst-case error for quantum processes. Though there is no simple way to calculate the diamond distance in general, it can be efficiently evaluated by the methods of semidefinite programming \cite{W2009, BT2010, W2013}.



We shall analyze how the quantum code affects two measures of the coherent error channels given by Eq. (\ref{noiseM}). In this case, both pre-error channel (given by Eq. (\ref{noiseM})) and post-error channel (given by Eq. (\ref{postnoise})) can be described below in terms of Pauli-Liouville representation.

\begin{equation}
\mathcal{E}= \left[\begin{array}{cccc}
1& 0 &0&0 \\
0      & x & 0&0\\
0 & 0& y&-u\\
0&0&u&y
\end{array}\right] 
\end{equation}

Using the formula given in Eq. (\ref{fidelity}), we immediately obtain the average gate infidelity
\begin{eqnarray}
r(\mathcal{E})=\frac{1}{6}(3-x-2y). \label{infidelity_r} \label{infidelityf}
\end{eqnarray}

Enlightened by the method of semidefinite programming employed in \cite{W2009, BT2010, W2013, MGE2012}, we can derive the diamond distance formula shown below
\begin{eqnarray}
\mathrm{D}_\diamond(\mathcal{E})=\frac{1}{4}(1-x)+\frac{1}{2}\sqrt{(1-y)^2+u^2}. \label{diamond_d} \label{diamondf}
\end{eqnarray}
The detailed deriving of this formula is deferred to Appendix C.

The two measurements of the pre-error channel $\mathcal{N}^{(1)}$ are shown below:

\begin{eqnarray}
\!\!r(\mathcal{N}^{(1)})
\!&=&\frac{1}{3}[1-(1-2q)\cos (\epsilon)]\approx \frac{2}{3}q+\frac{1}{6}\epsilon^2 \label{}\\
\!\!\mathrm{D}_\diamond(\mathcal{N}^{(1)})
\!&=&\frac{1}{2}\sqrt{1+(1-2q)^2-2(1-2q)\cos (\epsilon)}\\
\!&\approx& \sqrt{q^2+\frac{\epsilon^2}{4}} \label{diamond_r}
\end{eqnarray}

It is noted that when the rotation error dominates, $|\epsilon|\gg q$, we have $\mathrm{D}_\diamond(\mathcal{N}^{(1)})\approx \sqrt{\frac{3}{2}r(\mathcal{N}^{(1)})}$, this finding is similar to a result given in \cite{KLDF2015}.

The two measurements of the post-error channel $\Omega^{\mathrm{Five}}(\mathcal{N}^{(1)})$ are shown below:

\begin{eqnarray}
\!\!r(\Omega^{\mathrm{Five}}(\mathcal{N}^{(1)}))
\!&=&\frac{1}{24}(13-10y-10y^2+5y^4+2y^5) \nonumber \label{}\\
\!\!\mathrm{D}_\diamond(\Omega^{\mathrm{Five}}(\mathcal{N}^{(1)}))
\!&=&\frac{5}{16}(1-2y^2+y^4)\\
\!&+& \frac{1}{8}\sqrt{(4-5y+y^5)^2+u^{10}}\label{}
\end{eqnarray}
\noindent where $y=(1-2q)\cos (\epsilon)$ and $u=(1-2q)\sin (\epsilon)$.

It is noted that when the rotation error dominates, $|\epsilon|\gg q$, the error channel becomes coherent, we have $r(\Omega^{\mathrm{Five}}(\mathcal{N}^{(1)}))\approx \frac{5}{12}\epsilon^4$ and $\mathrm{D}_\diamond(\Omega^{\mathrm{Five}}(\mathcal{N}^{(1)}))\approx \frac{5}{8}\epsilon^4$. Remarkably, $\mathrm{D}_\diamond(\Omega^{\mathrm{Five}}(\mathcal{N}^{(1)}))\approx 10[\mathrm{D}_\diamond(\mathcal{N}^{(1)})]^4$, our result sharpens a general conclusion about the upper bound on the diamond distance after error correction for unitary channels and stabilizer codes obtained by \cite{HDF2019}, in which the upper bound is considered to be the third power of the diamond distance of a error channel at the physical level. For the opposite extreme case when the stochastic bit-flip probability dominates, $q\gg |\epsilon|$, the error channel becomes a pure Pauli channel, we have $r(\Omega^{\mathrm{Five}}(\mathcal{N}^{(1)}))\approx \frac{20}{3}q^2$ and $\mathrm{D}_\diamond(\Omega^{\mathrm{Five}}(\mathcal{N}^{(1)}))\approx 10q^2$. In each of the two cases mentioned above, we can find that the quantum error correction code can greatly reduce the strength of error.

\vskip 0.2in
\section{Appendix}

\subsection{\bf Coding map of the five-qubit code with the symmetric decoder}

For the five-qubit code with the symmetric decoder, the stabilizer, $S=\{S_k\}=\langle XZZXI, IXZZX, XIXZZ, ZXIXZ  \rangle$, the logical Pauli operators are $\bar{I}=IIIII$, $\bar{X}=XXXXX$, $\bar{Z}=ZZZZZ$, and $\bar{Y}=i\bar{X}\bar{Z}=YYYYY$, the recovery operator $R=\{IIIII, X_j, Y_j, Z_j:j=1,...,5\}$ where $\sigma_j$ stands for the tensor product of five Pauli matrices or the identity matrix with the $j$th component being $\sigma$ and the rest components are all identity matrix $I$. It is noted that this recovery operator is called the symmetric decoder for the code, which associates each syndrome to a unique weight-one Pauli operator. Therefore, all weight-one Pauli operators are corrected.

The encoding operation $\mathcal{E}$ and the decoding operation $\mathcal{D}$ are described as follows:

$E_{I}=\frac{1}{2|S|}\sum_{S_k\in S}S_k=\frac{1}{32}(IIIII+XZZXI)\times (IIIII+IXZZX)\times (IIIII+XIXZZ)\times (IIIII+ZXIXZ)$

$E_{\sigma}=E_I\bar{\sigma}$ where $\sigma\in\{X,Y, Z\}$

Since $D_{\sigma}=\frac{1}{|S|}\sum_{k,j}R^{\dagger}_jS_k\bar{\sigma}R_j=\frac{1}{|S|}\sum_kf_{k\sigma}S_k\bar{\sigma}$ where $f_{k\sigma}=\sum_j\eta(S_k, R_j)\eta(R_j,\bar{\sigma})$, we show explicit expressions for each $D_{\sigma}$:

\begin{widetext}
$D_{I}=IIIII$

\begin{eqnarray}
\!\!\!D_{X}
\!\!&=&-\frac{1}{4}(XXXXX+IYYIX+XIYYI+IXIYY+YIXIY+IZXZI+XYZZY+ZZYXY+IIZXZ \nonumber\\
\!\!&+&YXYZZ+ZIIZX+XZIIZ+ZYXYZ+YZZYX+ZXZII+YYIXI) \nonumber\\
\!\!\!D_{Y}
\!\!&=&-\frac{1}{4}(YYYYY+ZXXZY+YZXXZ+ZYZXX+XZYZX+ZIYIZ+YXIIX+IIXYX+ZZIYI \nonumber\\
\!\!&+&XYXII+IZZIY+YIZZI+IXYXI+XIIXY+IYIZZ+XXZYZ) \nonumber\\
\!\!\!D_{Z}
\!\!&=&-\frac{1}{4}(ZZZZZ+YIIYZ+ZYIIY+YZYII+IYZYI+YXZXY+ZIXXI+XXIZI+YYXZX \nonumber\\
\!\!&+&IZIXX+XYYXZ+ZXYYX+XIZIX+IXXIZ+XZXYY+IIYZY)
\end{eqnarray}

\end{widetext}
By the formulas given in Eqs. (\ref{unitalNC}) and (\ref{echannel}), we can obtain the entries of the process matrix of the coding map as shown below:

$G_{II}=1$, $G_{IX}=G_{IY}=G_{IZ}=0$

$G_{XI}=G_{YI}=G_{ZI}=0$

\begin{widetext}
$G_{XX}=h_x(N_{XX}, N_{XY}, N_{XZ}, N_{YX}, N_{YY}, N_{YZ}, N_{ZX}, N_{ZY}, N_{ZZ})=-\frac{N_{XX}}{4}(N_{XX}^4+5N_{XY}^2N_{XZ}^2-5 N_{YY}^2-5N_{ZZ}^2+5N_{YX}^2N_{ZX}^2+5N_{YY}^2N_{ZZ}^2)-\frac{5}{2}(N_{XY}N_{YX}N_{YZ}N_{ZY}N_{ZZ}+N_{XZ}N_{YY}N_{YZ}N_{ZX}N_{ZY})$

$G_{XY}=h_y(N_{XX}, N_{XY}, N_{XZ}, N_{YX}, N_{YY}, N_{YZ}, N_{ZX}, N_{ZY}, N_{ZZ})=-\frac{N_{XY}}{4}(N_{XY}^4+5N_{XX}^2N_{XZ}^2-5 N_{YZ}^2-5N_{ZX}^2+5N_{YY}^2N_{ZY}^2+5N_{YZ}^2N_{ZX}^2)-\frac{5}{2}(+N_{XX}N_{YX}N_{YZ}N_{ZY}N_{ZZ}+N_{XZ}N_{YX}N_{YY}N_{ZX}N_{ZZ})$

$G_{XZ}=h_z(N_{XX}, N_{XY}, N_{XZ}, N_{YX}, N_{YY}, N_{YZ}, N_{ZX}, N_{ZY}, N_{ZZ})=-\frac{N_{XZ}}{4}(N_{XZ}^4+5N_{XX}^2N_{XY}^2-5N_{YX}^2+5N_{YX}^2N_{ZY}^2+5N_{YZ}^2N_{ZZ}^2-5N_{ZY}^2)-\frac{5}{2}(N_{XX}N_{YY}N_{YZ}N_{ZX}N_{ZY}+N_{XY}N_{YX}N_{YY}N_{ZX}N_{ZZ})$

$G_{YX}=h_x(N_{YX}, N_{YZ}, N_{YY}, N_{XX}, N_{XZ}, N_{XY}, N_{ZX}, N_{ZZ}, N_{ZY})=-\frac{N_{YX}}{4}(N_{YX}^4+5N_{XX}^2N_{ZX}^2-5N_{XZ}^2+5N_{XZ}^2N_{ZY}^2+5N_{YY}^2N_{YZ}^2-5N_{ZY}^2)-\frac{5}{2}(N_{XX}N_{XY}N_{YZ}N_{ZY}N_{ZZ}+N_{XY}N_{XZ}N_{YY}N_{ZX}N_{ZZ})$

$G_{YY}=h_y(N_{YZ}, N_{YY}, N_{YX}, N_{XZ}, N_{XY}, N_{XX}, N_{ZZ}, N_{ZY}, N_{ZX})=-\frac{N_{YY}}{4}(N_{YY}^4+5N_{XX}^2N_{ZZ}^2-5 N_{XX}^2+5N_{XY}^2N_{ZY}^2+5N_{YX}^2N_{YZ}^2-5N_{ZZ}^2)-\frac{5}{2}(+N_{XX}N_{XZ}N_{YZ}N_{ZX}N_{ZY}+N_{XY}N_{XZ}N_{YX}N_{ZX}N_{ZZ})$

$G_{YZ}=h_z(N_{YY}, N_{YX}, N_{YZ}, N_{XY}, N_{XX}, N_{XZ}, N_{ZY}, N_{ZX}, N_{ZZ})=-\frac{N_{YZ}}{4}(N_{YZ}^4+5N_{XY}^2N_{ZX}^2-5N_{XY}^2+5N_{XZ}^2N_{ZZ}^2+5N_{YX}^2N_{YY}^2-5N_{ZX}^2)-\frac{5}{2}(N_{XX}N_{XZ}N_{YY}N_{ZX}N_{ZY}+N_{XX}N_{XY}N_{YX}N_{ZY}N_{ZZ})$

$G_{ZX}=h_x(N_{ZX}, N_{ZZ}, N_{ZY}, N_{YX}, N_{YZ}, N_{YY}, N_{XX}, N_{XZ}, N_{XY})=-\frac{N_{ZX}}{4}(N_{ZX}^4+5N_{XX}^2N_{YX}^2-5N_{XY}^2+5N_{XY}^2N_{YZ}^2-5N_{YZ}^2+5N_{ZY}^2N_{ZZ}^2)-\frac{5}{2}(N_{XX}N_{XZ}N_{YY}N_{YZ}N_{ZY}+N_{XY}N_{XZ}N_{YX}N_{YY}N_{ZZ})$

$G_{ZY}=h_y(N_{ZZ}, N_{ZY}, N_{ZX}, N_{YZ}, N_{YY}, N_{YX}, N_{XZ}, N_{XY}, N_{XX})=-\frac{N_{ZY}}{4}(N_{ZY}^4+5N_{XY}^2N_{YY}^2-5 N_{XZ}^2+5N_{XZ}^2N_{YX}^2-5N_{YX}^2+5N_{ZX}^2N_{ZZ}^2)-\frac{5}{2}(+N_{XX}N_{XY}N_{YX}N_{YZ}N_{ZZ}+N_{XX}N_{XZ}N_{YY}N_{YZ}N_{ZX})$

$G_{ZZ}=h_z(N_{ZY}, N_{ZX}, N_{ZZ}, N_{YY}, N_{YX}, N_{YZ}, N_{XY}, N_{XX}, N_{XZ})=-\frac{N_{ZZ}}{4}(N_{ZZ}^4+5N_{XX}^2N_{YY}^2-5N_{XX}^2+5N_{XZ}^2N_{YZ}^2-5N_{YY}^2+5N_{ZX}^2N_{ZY}^2)-\frac{5}{2}(N_{XX}N_{XY}N_{YX}N_{YZ}N_{ZY}+N_{XY}N_{XZ}N_{YX}N_{YY}N_{ZX})$

\vskip 0.2in
$h_x(x_1, x_2, x_3, y_1, y_2, y_3, z_1, z_2, z_3)=-\frac{x_1}{4}(x_1^4+5x_2^2x_3^2+5y_1^2z_1^2+5y_2^2z_3^2)-\frac{5}{2}y_3z_2(x_2y_1z_3+x_3y_2z_1)+\frac{5x_1}{4}(y_2^2+z_3^2)$

$h_y(x_1, x_2, x_3, y_1, y_2, y_3, z_1, z_2, z_3)=-\frac{x_2}{4}(x_2^4+5x_1^2x_3^2+5y_2^2z_2^2+5y_3^2z_1^2)-\frac{5}{2}y_1z_3(x_1y_3z_2+x_3y_2z_1)+\frac{5x_2}{4}(y_3^2+z_1^2)$

$h_z(x_1, x_2, x_3, y_1, y_2, y_3, z_1, z_2, z_3)=-\frac{x_3}{4}(x_3^4+5x_1^2x_2^2+5y_1^2z_2^2+5y_3^2z_3^2)-\frac{5}{2}y_2z_1(x_1y_3z_2+x_2y_1z_3)+\frac{5x_3}{4}(y_1^2+z_2^2)$

\end{widetext}
\subsection{ The convergence of the concatenated five-qubit code}

By the aforesaid analysis presented in III C, it suffices to study the dynamical system given by $f=(f_1, f_2): (x,y)\mapsto (f_1(x,y), f_2(x,y))$ where $f_1(x,y)=-\frac{x}{4}(x^4+5y^4-10y^2)$, $f_2(x,y)=-\frac{y}{4}(y^4+5x^2y^2-5x^2-5y^2)$, and $(x,y)\in \Delta\subset \mathbb{R}^2$. The total derivative

\begin{equation}
\mathrm{D}f(x,y)= \left[\begin{array}{cccc}
\frac{\partial f_1(x,y)}{\partial x}&\frac{\partial f_1(x,y)}{\partial y}\\
\frac{\partial f_2(x,y)}{\partial x}&\frac{\partial f_2(x,y)}{\partial y}
\end{array}\right] \label{},
\end{equation}
where $\left\{
\begin{array}{l}
\frac{\partial f_1(x,y)}{\partial x}=-\frac{5}{4}(x^4+y^4)+\frac{5}{2}y^2\\
\frac{\partial f_1(x,y)}{\partial y}=5xy(1-y^2)\\
\frac{\partial f_2(x,y)}{\partial x}=\frac{5}{2}xy(1-y^2)\\
\frac{\partial f_2(x,y)}{\partial y}=-\frac{5}{4}y^2(-3+3x^2+y^2)+\frac{5}{4}x^2
\end{array}\right.$

Using the formula of matrix 2-norm, given by $||A||_2=\sqrt{\rho(A^{\star}A)}$ where $\rho(A^{\star}A)$ is the largest eigenvalue of $A^{\star}A$, we can 
derive the norm $||\mathrm{D}f(x,y)||=\frac{5}{8}\sqrt{a+2\sqrt{b}}$ where $a$ and $b$ are given below:


\begin{widetext}
\begin{eqnarray}
\!\!a
\!&=&\!2(x^4+y^4-2y^2)^2+2(-3y^2+3x^2y^2+y^4-x^2)^2+32(xy-xy^3)^2+8(xy-xy^3)^2 \nonumber\\
\!\!b
\!&=&\!(x^8-x^4-7x^4y^4-18x^2y^6+2x^4y^2+44x^2y^4+2y^6-5y^4-18x^2y^2)^2\nonumber\\
\!&+&\!16(3xy^7-10xy^5+2x^5y^3+3x^3y^5-2x^5y-4x^3y^3+x^3y+7xy^3)^2
\end{eqnarray}


\begin{table}[h]
\caption{\mbox{\footnotesize{The approximate extreme values of $||\mathrm{D}f||$ on the circle of radius $r$ centered at the point $(1,1)$}}} 
\centering  
\begin{tabular}{c c c ccccc} 
\hline\hline                        
$r$ &.02 &.04 & .06 &.07 & \color{green}.072 &.073 & .08  \\ [0.5ex] 
\hline                  
$\mathrm{max} ||\mathrm{D}f(x,y)||$ & .245& .520& .812 &.964&\color{green}.995&1.010 &1.118  \\ 
\hline 
$\mathrm{min} ||\mathrm{D}f(x,y)||$ & .101& .202& .283 &.332&.346 &.361 &.387 \\ 
\hline\hline
\end{tabular}
\label{table:nonlin} 
\end{table}

\end{widetext}
\vskip 0.2in

\subsection{Diamond distance of the related channels}

We will use the semidefinite programming to evaluate the Diamond distance of the related channels. For $\Delta=I-\mathcal{E}$, according to \cite{W2009}, the following pair of primal and dual semidefinite programming has an optimal value of $\mathrm{D}_{\diamond}(\mathcal{E})=\frac{1}{2}||\Delta||_{\diamond}$: 

Primal problem: Maximize $\langle J(\Delta), W \rangle$ subject to $W\le \mathbb{I}_d\otimes \rho$, $W\in \mathrm{Pos}(L(\mathbb{C}^d\otimes \mathbb{C}^d))$, and $\rho\in D(L(\mathbb{C}^d))$.

Dual problem: Minimize $||\mathrm{Tr}_2(Z)||_{\infty}$ subject to $Z\ge J(\Delta)$, $z\in \mathrm{Pos}(L(\mathbb{C}^d\otimes \mathbb{C}^d))$,

\noindent where $J(\Delta)\coloneqq \sum_{i,j=1}^{d}|i\rangle\langle j|\otimes \Delta(|i\rangle\langle j|)$ is the Choi matrix of $\Delta$ \cite{Choi1975, MGE2012, KLDF2015}, $\langle X,Y\rangle=\mathrm{Tr}(X^{\dagger}Y)$ is the Hilbert-Schmidt inner product of the matrices $X$ and $Y$, $\mathrm{Pos}(L(\mathbb{C}^d\otimes \mathbb{C}^d))$ denotes the cone of positive semidefinite operators acting on the system $\mathbb{C}^d\otimes \mathbb{C}^d$, and $ \mathrm{Tr}_2(X)$ is the partial trace of $X$ over the second subsystem. Finally, $||X||_{\infty}$ denotes the operator norm of $X$,
which is the maximum eigenvalue of $X$(if $X\ge 0$). 

Let $\Pi_+$ be the projector onto the eigenspace of $J(\Delta)$ with positive eigenvalues, then $\rho=\frac{1}{d}I$ $W=\frac{1}{d}\Pi_+$ are valid primal feasible values and $Z=\Pi_+J(\Delta)\Pi_+$ is a dual feasible value. It is noteworthy that \cite{MGE2012} introduced these feasible points to evaluate the diamond distance of the noise channels. We will adopt these feasible points or similar values to calculate diamond distances of the channels in question.

{\bf The diamond distance of the coherent error}\,\,

Supposed that the Pauli-Liouville representation of an operator $\mathcal{E}$ is given below (this will always be the case for us),  we first compute the Choi matrix $J(\Delta)$ for $\Delta=I-\mathcal{E}$.

\begin{widetext}
\begin{equation}
\mathcal{E}= \left[\begin{array}{cccc}
1& 0 &0&0 \\
0      & x& 0&0\\
0 & 0& y&-u\\
0&0&u&y
\end{array}\right] \label{},
J(\Delta)= \left[\begin{array}{cccc}
\frac{1}{2}-\frac{1}{2}y& -i\frac{1}{2}u&-i\frac{1}{2}u&1-\frac{1}{2}x-\frac{1}{2}y\\
i\frac{1}{2}u& -\frac{1}{2}+\frac{1}{2}y&-\frac{1}{2}x+\frac{1}{2}y&i\frac{1}{2}u\\
i\frac{1}{2}u& -\frac{1}{2}x+\frac{1}{2}y&-\frac{1}{2}+\frac{1}{2}y&i\frac{1}{2}u \\
1-\frac{1}{2}x-\frac{1}{2}y& -i\frac{1}{2}u&-i\frac{1}{2}u&\frac{1}{2}-\frac{1}{2}y
\end{array}\right] \label{}
\end{equation}
\end{widetext}
With direct calculation, we obtain four eigenvalues of $J(\Delta)$ and a unital eigenvector of $J(\Delta)$ corresponding to positive eigenvalue $\lambda_1$ as follows: 

\begin{eqnarray}
\!\!\lambda_{1,2}
\!&=&\!\frac{1}{2}(1-x)\pm \sqrt{(1-y)^2+u^2}, \lambda_{3,4}=-\frac{1}{2}+\frac{1}{2}x \nonumber\\
\!\!u_1
\!&=&\![\frac{b}{a}, \frac{c}{a}, \frac{c}{a}, \frac{b}{a}]^T \label{+eigenvalue}
\end{eqnarray}

\noindent where $a=\sqrt{2u^2+2[\lambda_1-(\frac{3}{2}-\frac{1}{2}x-y)]^2}$, $b=-iu$, and  

\noindent $c=\lambda_1-(\frac{3}{2}-\frac{1}{2}x-y)$.



According to the spectral theorem for Hermitian matrices, there exists a unitay matrix $U$ and a diagonal matrix D such that $J(\Delta)=UDU^{\dagger}$ where $D$ has the above four (real valued ) eigenvalues as its diagonal elements. Suppose that $U=[u_1 ,u_2, u_3, u_4]$, then $u_1$, $u_2$, $u_3$ and $u_4$ are orthonormal. It is noted that $\{|\psi_1\rangle \coloneqq U|00\rangle,|\psi_2\rangle \coloneqq U|01\rangle,|\psi_3\rangle \coloneqq U|10\rangle, |\psi_4\rangle \coloneqq U|11\rangle \}$ form an orthonormal basis for the space $\mathbb{C}^2\otimes\mathbb{C}^2$, we have that $J(\Delta)$ is the diagonal matrix $D$ when written in the aforesaid orthonormal basis, then $\Pi_+$ would be the projector onto the eigenspace with the positive eigenvalue given in Eq (\ref{+eigenvalue}).

$|\psi_1\rangle \coloneqq U|00\rangle=u_1^T[|00\rangle, |01\rangle, |10\rangle, |11\rangle]^T$

For the primal problem let $W=\frac{1}{2}\Pi_+$ and $\rho=\frac{1}{2}\mathbb{I}_2$, then $\langle J(\Delta), W\rangle=\frac{1}{2}\lambda_1$.

For the dual problem take $Z=\Pi_+J(\Delta)\Pi_+$ which is just $\lambda_1|\psi_1\rangle\langle \psi_1|$ and note that $Z\ge J(\Delta)$. Moreover $\mathrm{Tr}_2Z=\frac{1}{2}\lambda_1\mathbb{I}_2$ and so $||\mathrm{Tr}_2Z||_{\infty}=\frac{1}{2}\lambda_1$

Therefore we obtain 

\begin{eqnarray}
\mathrm{D}_\diamond(\mathcal{E})=\frac{1}{4}(1-x)+\frac{1}{2}\sqrt{(1-y)^2+u^2}. \label{diamond_r}
\end{eqnarray}

\begin{eqnarray}
\mathrm{D}_\diamond(\mathcal{N}^{(1)})=\frac{1}{2}\sqrt{1+(1-2q)^2-2(1-2q)\cos (\epsilon)}. \label{diamond_r}
\end{eqnarray}
 
\vskip 0.2in

\section{ACKNOWLEDGMENTS}

This work was supported by NSF award number 1714261.


\smallskip

\begin{thebibliography}{9}

\bibitem{NC2000}

M. Nielsen and I. Chuang, {\it Quantum Computation and Quantum Information} (Cambridge
University Press, Cambridge, 2000).


\bibitem{Shor1995}

P.W. Shor. Scheme for reducing decoherence in quantum computer memory. Phys. Rev. A., 52:R2493, 1995.

\bibitem{Steane1996a}

Steane, A.M. (1996). Error correcting codes in quantum theory. Physical Review Letters, 77,793.

\bibitem{Bennett1996}
Bennett, C.H., DiVincenzo, D.P., Smolin, J.A. and Wootters, W.K. (1996). Mixed state-entanglement and quantum error correction. Physical Review A, 54, 3824.

\bibitem{Laflamme1996}
Laflamme, R., Miquel, C., Paz, J.-P. and Zurek, W.H. (1996). Perfect quantum error correction
code. Physical Review Letters, 77, 198. 

\bibitem{G1997}
Daniel Gottesman, Stabilizer Codes and Quantum Error Correction, quant-ph/9705052, Caltech Ph.D. thesis.

\bibitem{FM2012}

Austin G. Fowler, Matteo Mariantoni, John M. Martinis, and Andrew N. Cleland, Surface codes: Towards practical large-scale quantum computation, Phys. Rev. A 86, 032324 (2012). 

\bibitem{DNM2013}

Simon J. Devitt, Kae Nemoto, William J. Munro, Quantum Error Correction for Beginners, Rep. Prog. Phys. 76 (2013) 076001.

\bibitem{T2015}
Barbara M. Terhal, Quantum Error Correction for Quantum Memories, Rev. Mod. Phys. 87, 307 (2015).

\bibitem{Brun2019}

Todd A. Brun, Quantum Error Correction, arXiv:1910.03672 (2019).


\bibitem{GS2016}

Mauricio Guti$\acute{e}$rrez, Conor Smith, Livia Lulushi, Smitha Janardan, and Kenneth R. Brown, Errors and pseudo$\-$thresholds for incoherent and coherent noise, Phys. Rev. A 94, 042338 (2016), arXiv:1605.03604.

\bibitem{DP2017}

Andrew S. Darmawan and David Poulin,Tensor$\-$Network Simulations of the Surface Code under Realistic Noise, Physical Review Letters 119, 040502 (2017)

\bibitem{BEKP2017}

Sergey Bravyi, Matthias Englbrecht, Robert Koenig, and Nolan Peard, Correcting coherent errors with surface codes, npj Quantum Information, vol. 4, no. 55 (2018).


\bibitem{GD2018}
Daniel Greenbaum and Zachary Dutton, Modeling coherent errors in quantum error correction, Quantum Sci.
Technol. 3, 015007 (2018), arXiv:1612.03908.


\bibitem{BW2018}

Stefanie Beale, Joel Wallman, Mauricio Gutierrez, Kenneth R. Brown, and Ray Laflamme, Quantum error correction decoheres noise, Phys. Rev. Lett. 121, 190501 (2018).

\bibitem{IP2020}

Joseph K Iverson and John Preskill, Coherence in logical quantum channels, 2020 New J. Phys. 22 073066.

\bibitem{CXB2020}

Zhenyu Cai, Xiaosi Xu and Simon C. Benjamin, Mitigating coherent noise using Pauli conjugation, npj Quantum Information volume 6, Article number: 17 (2020). 

\bibitem{BW2021}

Stefanie J. Beale, Joel J. Wallman, Efficiently computing logical noise in quantum error correcting codes, Phys. Rev. A 103, 062404 (2021).

\bibitem{GY2022}

Ming Gong, Xiao Yuan et al., Experimental exploration of five-qubit quantum error-correcting code with superconducting qubits , National Science Review, Volume 9, Issue 1, January 2022, nwab011.


\bibitem{RDM2002}
 B. Rahn, A. C. Doherty, and H. Mabuchi, Exact perfomance of concatenated quantum codes,  Phys. Rev. A, vol. 66, no. 032 304, 2002.


\bibitem{FKSS2006}

J. Fern, J. Kempe, S. N. Simic, and S. Sastry, Generalized performance of concatenated quantum codes-a dynamical systems approach, IEEE Transactions on Automatic Control 51, 448 (2006), quant-ph/0409084.


\bibitem{CW2017}

Christopher Chamberland, Joel Wallman, Stefanie Beale, and Raymond Laflamme, Hard decoding algorithm
for optimizing thresholds under general markovian noise, Phys. Rev. A 95, 042332 (2017), arXiv:1612.02830.

\bibitem{HDF2019}
E. Huang, A. C. Doherty, and S. Flammia,Performance of quantum error correction with coherent errors, Phys. Rev. A 99, 022313 (2019)



\bibitem{KLR2008}

E. Knill, D. Leibfried, R. Reichle, J. Britton, R. B. Blakestad, J. D. Jost, C. Langer, R. Ozeri, S. Seidelin, and D. J. Wineland, Randomized benchmarking of quantum gates, Phys. Rev. A 77, 012307 (2008),
arXiv:0707.0963.

\bibitem{MGE2011}
Easwar Magesan, Jay M. Gambetta, and
Joseph Emerson, Scalable and Robust Randomized Benchmarking of Quantum Processes,
Physical Review Letters 106, 180504 (2011).

\bibitem{N2002}
 M. A. Nielsen, A simple formula for the average gate fidelity of a quantum dynamical operation, Phys. Lett. A 303, 249 (2002), arXiv:quantph/0205035.
 
\bibitem{HHH1999}
 M. Horodecki, P. Horodecki, and R. Horodecki, General teleportation channel, singlet fraction and quasi-distillation, Phys. Rev. A
60, 1888 (1999), quant-ph/9807091

\bibitem{KLDF2015}

Richard Kueng, David M. Long, Andrew C. Doherty, Steven T. Flammia, Comparing Experiments to the Fault-Tolerance Threshold, Phys. Rev. Lett. 117, 170502 (2016).

\bibitem{K1997}



A. Y. Kitaev, Quantum computations: algorithms and error correction, Russian Math. Surv. 52, 1191 (1997).


\bibitem{W2009}
J. Watrous, Semidefinite Programs for Completely Bounded Norms,  Theory of Computing 5, 217 (2009).

\bibitem{BT2010}
A. Ben-Aroya and A. Ta-Shma, On the complexity of approximating the diamond norm, Quant. Inf. Comp. 10, 77
(2010).



\bibitem{W2013}
J. Watrous, Simpler semidefinite programs for completely bounded norms, Chicago J. Theo. Comp. Sci. 2013, 1 (2013).

\bibitem{MGE2012}

 E. Magesan, J. M. Gambetta, and J. Emerson, Characterizing Quantum Gates via Randomized Benchmarking, Phys. Rev. A 85,
042311 (2012), arXiv:1109.6887.

\bibitem{Choi1975}

M. Choi, Completely positive linear maps on complex matrices, Lin. Alg. Appl. 285-290 (1975).


\end{thebibliography}
\end{document}